\documentclass[doublecol]{epl2} 
\usepackage{epsfig}
\newcommand{\be}{\begin{equation}}
\newcommand{\ee}{\end{equation}}
\renewcommand{\phi}{\varphi}

\title{Compressing nearly hard sphere fluids increases glass fragility}
\shorttitle{Compressing nearly hard sphere fluids increases glass fragility} 

\author{Ludovic Berthier\inst{1} \and Thomas A. Witten \inst{2}}
\shortauthor{L. Berthier and T. A. Witten}

\institute{                    
  \inst{1} Laboratoire des Collo{\"\i}des, Verres
 et Nanomat{\'e}riaux, UMR 5587, \\ Universit{\'e} Montpellier II and CNRS,
34095 Montpellier, France \\
  \inst{2} James Franck Institute and Department of Physics,
The University of Chicago, \\ 929 E. 57th Street, Chicago, Illinois 60637, USA
}


\abstract{We use molecular dynamics to investigate the glass transition
occuring at large volume fraction, $\phi$, and low temperature, $T$, 
in assemblies of soft repulsive particles. 
We find that equilibrium dynamics in the ($\phi$, $T$) plane
obey a form of dynamic scaling in the proximity of 
a critical point at $T=0$ and $\phi=\phi_0$,
which should correspond to the ideal glass transition
of hard spheres.  This glass point, `point $G$', 
is distinct from athermal jamming thresholds.
A remarkable consequence of scaling behaviour is that
the dynamics at fixed $\phi$ passes smoothly from that of a
strong glass to that of a very fragile glass as $\varphi$ increases
beyond $\varphi_0$. Correlations between fragility and 
various physical properties are explored.}

\begin{document}

\maketitle

\section{Introduction}

Structureless, hard, frictionless particles pass from a mobile to an 
immobile state with increasing density~\cite{rmpgrains}. 
Interacting particles and molecules 
in a glass-forming material
also pass from a 
mobile fluid state to an immobile glassy
state as temperature is reduced~\cite{reviewnature}. 
Much research in the last decade has 
been devoted to extracting a common 
geometric essence from these two classes of phenomena. One line of 
research approaches the threshold 
of immobilization or jamming via processes unrelated 
to thermal equilibrium~\cite{torquato2,pointJ,teitel,hatano}. 
Connections between these jamming transitions and those seen at positive 
temperature are suggested~\cite{liunagel}, but remain unclear.
In particular, the threshold density for jamming 
and that for immobilization at non-zero temperature
are considered identical by some researchers~\cite{ken,chaikin}, 
distinct for others~\cite{luca,zamponi,jorge}, ill-defined by some 
others~\cite{torquato2,torquato}. Direct measurements 
are not conclusive, because the location of the glass 
transition relies upon fitting and extrapolation~\cite{chaikin,luca}, 
while the jamming transition is not uniquely 
defined~\cite{torquato2,pointJ}.

The notion that temperature and density should have analogous 
effects on the glass transition has a long
history~\cite{free,andersen}, although quantitative evidence 
supporting these analogies is limited. 
Recent work studying the effect of pressure on the glass transition
showed that the dynamics of glass-formers is little 
affected by increasing the density, since 
a simple rescaling procedure collapses a broad range of 
dynamic data~\cite{casalini,alba}. This finding directly implies that 
the (isochoric) fragility~\cite{angell} of most glass-formers is independent 
of density, at least in the range currently explored  
by experiments. A second conclusion is that the glass transition
of molecular systems is mostly controlled by 
temperature, suggesting that the density-driven glass transition
of hard spheres might have a different nature.

In this article we study the relative influence of density 
and temperature on the glass transition using 
a model of soft repulsive particles~\cite{durian}. 
In the zero-temperature limit,
the model is equivalent to density-controlled 
hard spheres, while it resembles thermally driven 
dense fluids at large density and finite temperature.
Another  motivation to use compressible 
particles is to access densities beyond the hard sphere critical density
for kinetic arrest that are unreachable with the hard sphere potential. This 
approach has proven useful in the context of athermal 
jamming~\cite{pointJ,teitel,hatano}, but was not extended to thermal 
equilibrium before. At equilibrium, 
issues related to the possible protocol dependence of 
the results~\cite{comment} do not arise.  

\begin{figure}
\onefigure[width=8.5cm]{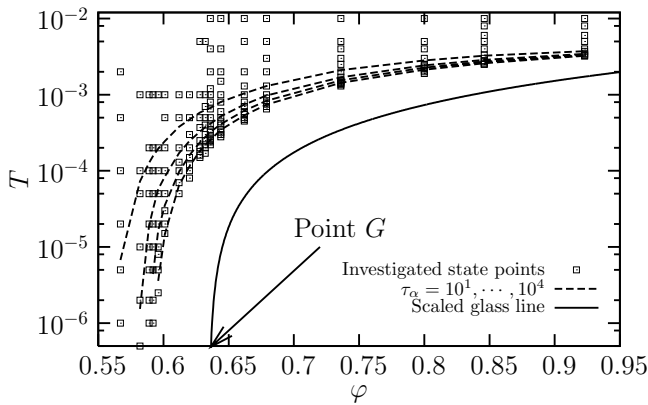}
\caption{\label{phase} 
Volume fraction, $\phi$, and temperature, $T$,  
phase diagram of elastic spheres with investigated state points. 
Four iso-relaxation time (dashed) lines 
are displayed, showing that $\tau_\alpha$ increases  
by increasing $\phi$ or decreasing $T$. 
Scaling permits accurate determination of 
the glass point $G$ at $T=0$ and $\phi_0=0.635$. The
glass transition (full) line is obtained 
assuming a specific form of the scaling function in eq.~(\ref{critical}).}
\end{figure}
  
Using computer simulations we have studied
the equilibrium dynamics of a three-dimensional 
assembly of soft repulsive particles
varying the volume fraction, $\phi$, and temperature, 
$T$ (see fig.~\ref{phase}). We have discovered a 
simple connection between density and temperature 
effects, from which several interesting results are deduced. 
Equilibrium dynamics obey critical scaling
in the proximity of a glass critical point, which we call
`point $G$', at $T=0$ and $\phi=\phi_0$, see fig.~\ref{phase}.
By approaching point $G$ along various equilibrium routes, we 
can determine the functional form of the dynamics 
and the location of the singularity with much less uncertainty than
is usually possible in glasses. 
In particular, we confirm 
the non-trivial density dependence of the relaxation time 
suggested by recent experiments on colloidal hard spheres~\cite{luca}. 
Scaling also implies that 
the evolution of the dynamics with temperature 
at fixed $\phi$ passes smoothly from that of a
strong glass to that of a very fragile glass as $\varphi$ increases
beyond $\varphi_0$. Compared to previous
numerical glass models with tunable fragility, the present model
does not require changing the composition of the 
liquid~\cite{coslovich}, or the curvature of space~\cite{sausset}, 
fragility varies over a much broader range~\cite{sri}, and provides
a new conceptual way to tune fragility.

\section{Model and methods}

We use Molecular Dynamics simulations~\cite{allen} to 
study a system composed of $N$ particles enclosed 
in a periodic cube of linear size $L$ and interacting through a 
pair-wise potential: $V(r_{ij}) = \epsilon (1- r_{ij}/\sigma_{ij})^2$ for
$r_{ij} < \sigma_{ij}$, $V(r_{ij})=0$ 
otherwise. The interparticle distance is $r_{ij} = |{\bf r}_i
-{\bf r}_j|$ and $\sigma_{ij} = (\sigma_i + \sigma_j)/2$,
where ${\bf r}_i$ and $\sigma_i$ are the position and diameter 
of particle $i$, respectively. We use system sizes between 
500 and 8000 particles, and report 
results for $N=1000$, for which no finite size effects 
are detected, within numerical accuracy.
We prevent crystallisation by using 
a 50:50 binary mixture of spheres of diameter ratio 
$1.4$~\cite{pointJ}.
The volume fraction is $\varphi = \frac{\pi N}{12 L^3} (1 
+ 1.4^3) \approx 0.98 \rho$, 
with $L$ expressed in units of the small particle 
diameter and $\rho=N/L$ the number density.
Up to volume fraction 
$\varphi = 0.846$ we detect no sign of crystallization at all 
studied temperatures; above $\phi = 0.924$ there was 
evidence of incipient crystallization at the lowest temperatures.
However, these crystalization effects occur well away from 
the region of interest around $\phi_0 = 0.635$.
We use $\epsilon$ as the energy unit, and 
$\sqrt{\sigma_2^2/\epsilon}$ as time unit, masses are set to unity.
All dynamical 
results are obtained at thermal equilibrium, which has been carefully 
controlled. When temperature is low and density is large, we are not
able to thermalize.
Crystallization and equilibrium issues determine the boundaries of the 
region investigated in the phase diagram of fig.~\ref{phase}.  

\section{Dynamic scaling at thermal equilibrium}

\begin{figure}
\psfig{file=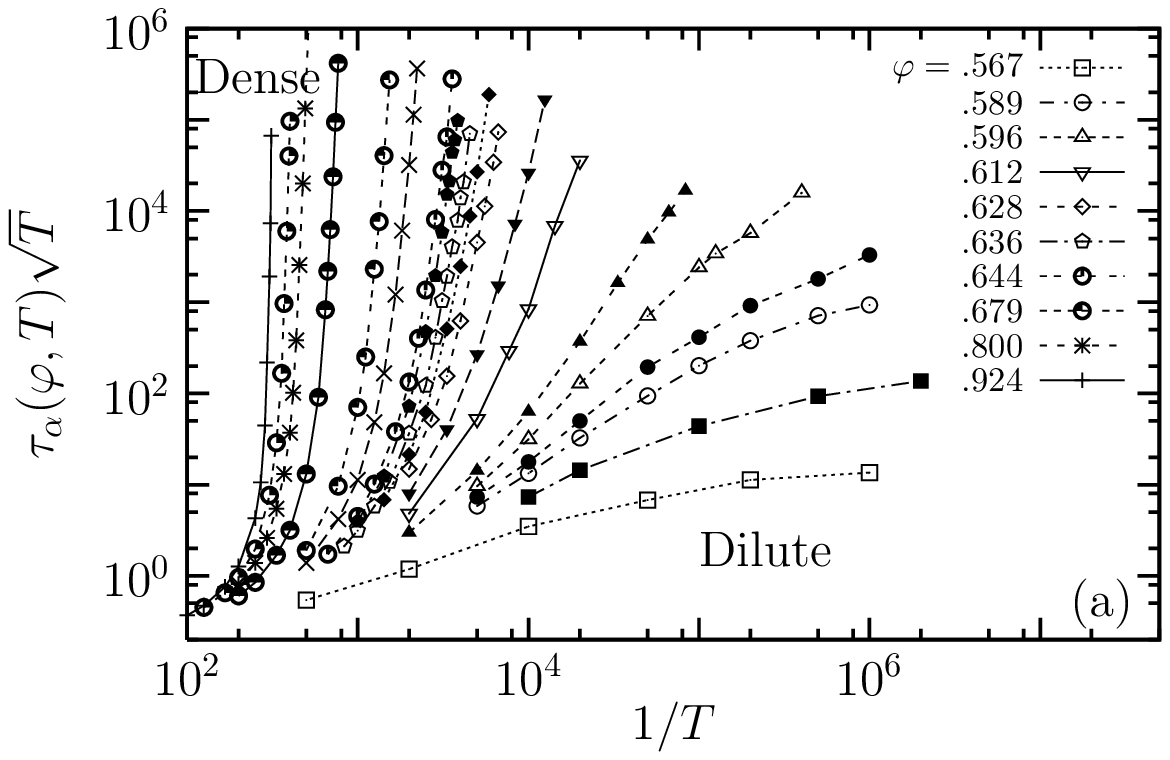,width=8.35cm}
\psfig{file=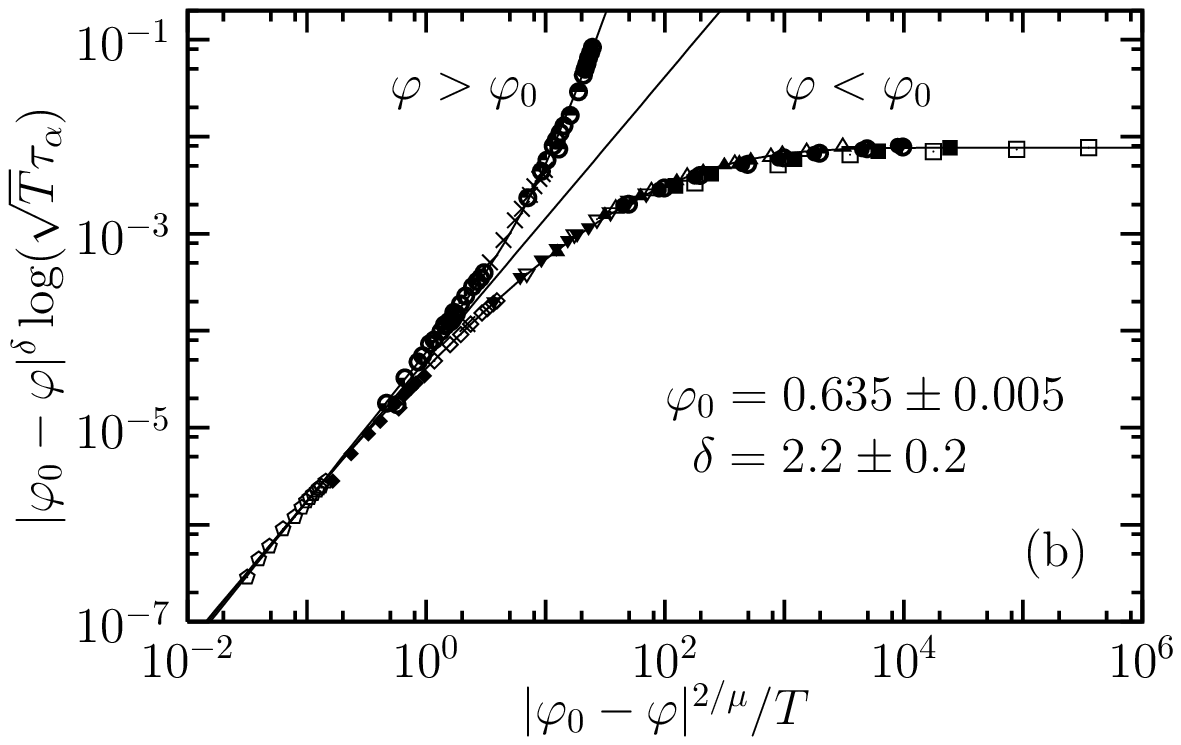,width=8.5cm}
\psfig{file=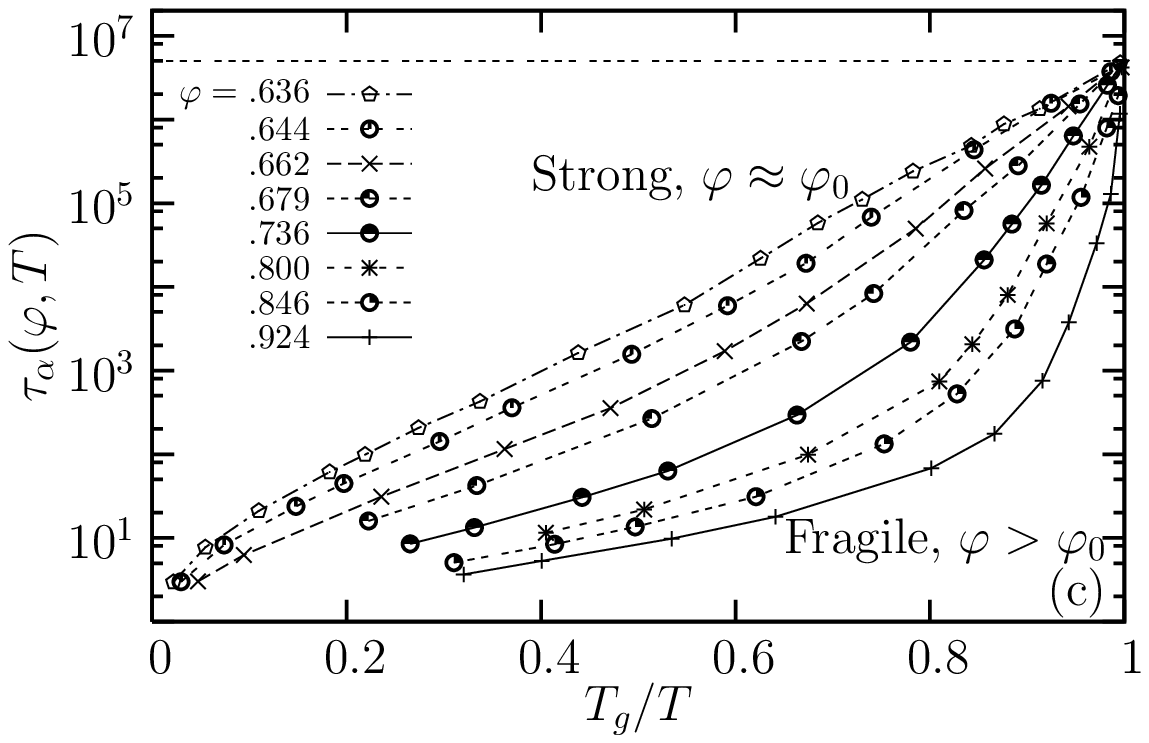,width=8.5cm}
\caption{\label{taualpha} 
(a) Relaxation timescales for all investigated state points.
We rescale $\tau_\alpha$ by $1/\sqrt{T}$ so that 
the $T \to 0$ limit coincides with hard spheres thermalized at $T=1$.
Not all volume fractions are labelled. 
(b) Collapse of the $\phi<\phi_0$ and $\phi>\phi_0$ data along the 
bottom and top branches, respectively using eq.~(\ref{critical})
and imposing $\mu=1.3$ from considering the potential energy.
(c) Arrhenius plot of the $\phi>\phi_0$ data, using 
the definition $\tau_\alpha(\phi,T_g)=5\cdot10^6$.}
\end{figure}

In fig.~\ref{taualpha}-a we report the evolution of the 
averaged relaxation timescale, $\tau_\alpha(\phi,T)$ 
for all investigated state points. 
We quantify  the microscopic dynamics through the 
self-part of the intermediate scattering function:
\be
F_s(q,t) = \frac{1}{N_b} \left\langle \sum_{j=1}^{N_b} e^{i
{\bf q} \cdot ({\bf r}_j(t)-{\bf r}_j(0))} \right\rangle,
\label{fseq}
\ee
where $\mathbf{q}$ is the scattering vector ($q = 6.1$,
close to the first diffraction peak) and
brackets indicate a thermal average.
We define $\tau_\alpha$ by $F_s(q,\tau_\alpha)=e^{-1}$,
and we arbitrarily choose to restrict the average in (\ref{fseq}) to the 
$N_b$ big particles. Data for $\tau_\alpha$
are normalized by $1/\sqrt{T}$, which is equivalent to 
renormalizing times by the `thermal' time $\frac{1}{v_{\rm MB}}$, 
where $v_{\rm MB}$ is the first moment of 
the Maxwell-Boltzmann distribution. 
Therefore, in the $T \to 0$ limit where particle overlaps
are energetically disfavoured, the dynamics of the elastic spheres 
coincide with that of hard spheres thermalized at $T=1$. 
We have verified this equivalence quantitatively 
by comparing our numerical results to the hard sphere 
studies presented in Ref.~\cite{luca}. 

The temperature evolution of $\tau_\alpha$ exhibits
two qualitatively distinct regimes.
For dilute systems, $\tau_\alpha$ first increases  
when $T$ decreases, but saturates when $T\to 0$ 
at a finite value corresponding to $\tau_\alpha^{\rm hs}(\phi)$, 
the relaxation time of the hard sphere fluid.
For dense systems, $\tau_\alpha$ seems 
to increase with no saturation, and we cannot reach equilibrium 
when $T$ becomes too small and the system becomes an unequilibrated glass
(see fig.~\ref{phase}).
The frontier between these two regimes must correspond 
to $\phi=\phi_0$, the density at which $\tau_\alpha^{\rm hs}(\phi)$
diverges.  By analogy with the {\it J}amming 
transition occuring at point $J$ along the $\phi$-axis
for athermal systems of spheres~\cite{pointJ}, we 
call `point $G$' the location of the {\it G}lass transition 
at $\phi_0$ along the $\phi$-axis for thermal systems.  

We now formulate the hypothesis that dynamics in the proximity 
of point $G$ obey scaling behaviour. We surmise that elastic
spheres at $T>0$ behave, in a statistical sense, 
similarly to hard spheres with a reduced 
diameter~\cite{free}.
Physically, we assume that at low $T$, 
overlaps between particles are very small, and so are 
corrections to hard sphere behaviour. 
Below, we justify why an accurate relation 
between density and temperature is: 
\be
\phi_{\rm eff}(\phi,T) \approx \phi - a T^{\mu/2},
\label{mapping}
\ee
where $a>0$ is a numerical prefactor, $\mu > 0$
a characteristic exponent, and $\phi_{\rm eff} (\phi,T) < \phi$ 
an effective volume fraction for the elastic sphere 
system at volume fraction $\phi$ and temperature $T$. 
Furthermore, we build upon a 
recent analysis of the dynamics of colloidal hard spheres~\cite{luca} 
and assume an exponential divergence for $\tau_\alpha^{\rm hs}$:
\be
\tau_\alpha^{\rm hs} (\varphi) \sim
\exp \left[ \frac{A}{(\varphi_{0} -
\varphi)^{\delta} } \right],
\label{vft} 
\ee
where $\delta \approx 2$ and $\phi_0 \approx 0.637$~\cite{luca}. 
Although the exponential divergence 
of $\tau_\alpha^{\rm hs}$ is unambiguous from hard sphere 
studies~\cite{luca}, the values of $\delta$
and $\phi_0$ remain subject to large uncertainty because  
the divergence must be extrapolated along a single path (increasing
$\phi$ at $T=0$), relatively far away from $\phi_0$. 

We analyze the results of fig.~\ref{taualpha}-a 
and approach point $G$ from multiple paths in the $(\phi,T)$ plane
to establish the robustness of eq.~(\ref{vft}). 
Combining (\ref{mapping})-(\ref{vft}) we suggest:
\be 
\tau_\alpha(\varphi,T) \sim \exp 
\left[ 
\frac{A}
{|\varphi_0-\varphi|^\delta} F_{\pm} \left( \frac{|\phi_0
-\phi|^{2/\mu}}{T} 
\right) \right], 
\label{critical} 
\ee 
where the scaling functions $F_\pm(x)$ apply to 
densities above/below $\phi_0$. 
We expect therefore that $F_-(x \to \infty) \to 1$ to recover the
hard sphere fluid limit, eq.~(\ref{vft}), when 
$T\to0$ and  $\phi < \phi_0$. 
Similarly, $F_+(x \to \infty) \to \infty$. Moreover, 
continuity of $\tau_\alpha$ at finite $T$ 
and $\phi=\phi_0$ implies $F_-(x\to 0) \sim F_+(x \to 0) 
\sim x^{\delta \mu / 2}$. 
Dynamic scaling was recently observed for athermal jamming 
transitions~\cite{teitel,hatano}, but the nature of the critical density 
and hard sphere divergence (algebraic instead of exponential)
were different from eq.~(\ref{critical}), 
while no physical interpretation of scaling
in terms of an effective hard sphere behaviour was offered.

The proposed scaling behaviour 
is confirmed in fig.~\ref{taualpha}-b for
data in the range $\phi \in [0.567, 0.736]$. 
To obtain the scaling plot, we fix $\mu =1.3$ (see below), 
and use $\phi_0$ and $\delta$ as free parameters
to collapse $|\phi_0 - \phi|^\delta \log \tau_\alpha$
against $|\phi_0-\phi|^{2/\mu}/T$. 
The best collapse is shown, but good results
are obtained for nearby values 
of $\phi_0$ and $\delta$, yielding error bars:
\be
\phi_0 = 0.635 \pm 0.005, \quad \delta = 2.2 \pm 0.2.
\ee 
Outside this range, the collapse quickly deteriorates. 
Note that $\delta=1$, often used to 
describe hard sphere data~\cite{chaikin}, 
is inconsistent with our results. Our scaling analysis thus lends 
crucial support to the conclusions of \cite{luca}. Of course, 
we cannot exclude that a different dynamic regime can be entered 
when relaxation timescales beyond reach of our numerical
capabilities are added to the analysis. 

\section{Glass fragility}

\begin{figure}
\psfig{file=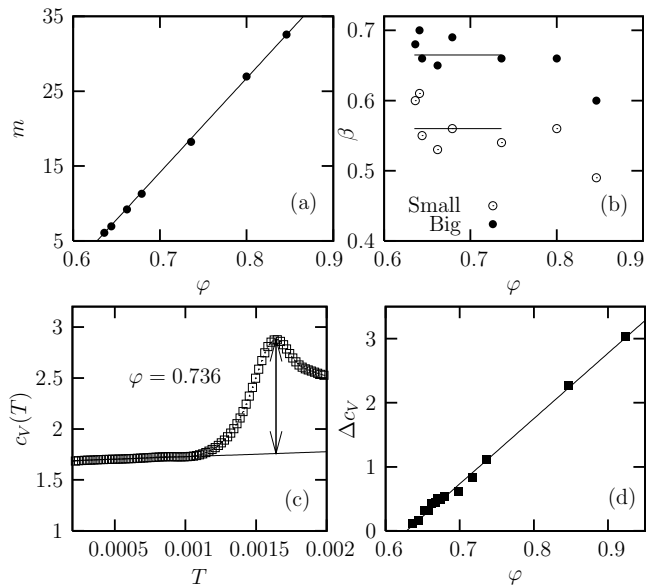,width=8.5cm}
\caption{\label{multi} (a) $\phi$ dependence of the steepness index,
eq.~(\ref{steepness}), 
measured from the data in fig.~\ref{taualpha}-c. The line 
is the linear fit predicted in eq.~(\ref{linear}).
(b) The stretching exponent describing the time decay of 
$F_s(q,t)$ for both species coincides with the one for hard spheres 
in the scaling regime of fig.~\ref{taualpha}-b, 
uncorrelated to fragility.
(c) Temperature dependence of the specific heat 
at $\phi=0.736$ obtained by heating the glass slowly. The full 
line fits the $T$-dependence of $c_V(T)$ below $T_g$, allowing
the definition of $\Delta c_V$, as shown by the vertical arrow.
(d) The $\phi$-dependence of $\Delta c_V$ (symbols) is well
described by a linear increase (full line), 
closely tracking the fragility in (a).}
\end{figure}

The scaling in eq.~(\ref{critical})
predicts the temperature 
dependence of $\tau_\alpha$ at $\phi_0$:
$\tau_\alpha(\varphi_0,T) \sim \exp 
\left( A / T^{\mu \delta/2} \right)$.
Since $\mu \delta /2 \approx 1.43$, this divergence
is slightly stronger than, 
but not very different from, the simple Arrhenius behaviour
observed for `strong' glass-formers~\cite{angell}. 
The divergence of the scaling function
$F_+(x)$ for large argument moreover implies that 
the temperature dependence of $\tau_\alpha$ for $\phi > \phi_0$
becomes steeper, making the materials
increasingly `fragile'~\cite{angell}. This is
vividly demonstrated in fig.~\ref{taualpha}-c, where
we conventionally rescale $T$ by $T_g$ defined as the temperature where 
$\tau_\alpha$ reaches an arbitrary value~\cite{angell}, 
$\log_{10} \tau_\alpha (\phi,T_g) = X_g$.
Such a large change of fragility
was not reported in a particle model before~\cite{coslovich,sausset,sri}.
Here, it directly results from the interplay between 
$\phi$ and $T$. We quantify fragility by the steepness
index~\cite{angell}: 
\be
m = \frac{\partial \log_{10} 
\tau_\alpha}{\partial (T_g/T)} {\bigg |}_{T_g},
\label{steepness}
\ee 
which increases
steadily when $\phi$ increases beyond $\phi_0$, see fig.~\ref{multi}-a.
The linear variation can be rationalized realizing that
an approximate expression for $F_+(x)$ in eq.~(\ref{critical})
can be obtained by pushing further 
our correspondance between elastic and hard spheres, and assuming:
$\tau_\alpha(\phi,T) \approx \tau_\alpha^{\rm hs}[\phi_{\rm eff}
(\phi,T)]$. This gives $F_+(x) \approx (a x^{-\mu/2} - 1)^{-\delta}$, from  
which we get:
\be
m (\phi_0 + \Delta \phi) \approx m_0 ( 1 + \alpha \Delta \phi ),
\label{linear}
\ee
where $m_0 = \mu \delta X_g/2$ and $\alpha = a (X_g/A)^{1/\delta}$.
Figure~\ref{multi}-a shows that this predicted linear behaviour 
is accurately obeyed over a seven-fold range of fragility $m$. 
A broader range of $m$ would be obtained for $X_g$ 
corresponding to the laboratory $T_g$. Multiplying
for instance $X_g$ by a factor 3 (from $X_g=5$ to $15$) 
in Eq.~(\ref{linear}) yields $m \in [18, 150]$, quite 
close to the experimental spectrum for the fragility
of glass-formers~\cite{angell}.

Additionally, the explicit, but approximate, expression for $F_+(x)$ 
predicts the location of a glass line, $T_0(\phi) = [
(\phi-\phi_0)/a^{1/\delta}]^{2/\mu}$, shown in fig.~\ref{phase}, 
in the spirit of \cite{andersen}. Unfortunately, 
our numerical data alone do not allow us to determine whether
$F_+(x)$ indeed diverges for a finite value of its argument.
Therefore, while the location of point $G$ is very much constrained
by our data, the existence of a finite temperature singularity 
for $\phi > \phi_0$ remains open.

Different glasses possess different
fragilities, but this variability is not satisfactorily understood. 
Instead, experimentalists often correlate
the properties of a glass to its 
fragility~\cite{angell}. Having a model with tunable fragility
allows theoretical exploration of the part 
of these correlations that is not due to the variety
of structures (e.g. networks, polymer chains) 
observed in real glass-formers.
We have studied two popular correlations:
the stretching of time correlation functions and 
the specific heat jump at $T_g$.
We fit the alpha-relaxation in $F_s(q,t)$ for both species
with a stretched exponential form: $F_s \sim
e^{-(t/\tau)^\beta}$. We find that $\beta$ is very weakly $T$-dependent,
and report its value for different $\phi$
in fig.~\ref{multi}-b. For $\phi$ within the scaling regime of 
fig.~\ref{taualpha}-b, the observed $\beta$'s for small and big 
particles are independent of compression, and hence of fragility, but
are consistent with those of hard spheres observed previously~\cite{luca}.
While some degree of correlation is often 
reported in experiments~\cite{reviewnature}, 
it is not very strong when data for polymers are discarded from the 
analysis~\cite{heuer}, in agreement with our findings.

We measure the potential energy,
$V_{\rm pot}(\phi,T) = 
\langle \sum_{i < j} V(r_{ij}) \rangle /N$, and
the specific heat, $c_V = d V_{\rm pot}/dT$. Mimicking experiments,
we obtain the jump in $c_V$ at the glass transition
by slowly cooling the system, at constant $\phi$, down to $T=0$
before re-heating  at the same rate. For all $\phi$,
we use very slow rates, $\frac{1}{T_g} \frac{d T}{ dt} 
\approx 3.10^{-7}$. The typical behaviour
of $c_V(T)$ upon heating is shown in fig.~\ref{multi}-c. 
As in experiments, a peak is observed when equilibrium
is recovered. We estimate $\Delta c_V$ as the difference between   
the peak height and the glass specific heat, although 
different estimates yield qualitatively similar results.
The $\phi$-dependence of $\Delta c_V$ is shown
in fig.~\ref{multi}-d, together with a linear fit. 
As suggested by experiments~\cite{reviewnature,angell}, we find a linear 
relation between fragility and specific 
heat. 

To qualitatively explain this observation we discuss
the behaviour of $V_{\rm pot}$. For a very dilute system, 
$V_{\rm pot}$ decreases rapidly as $T$ decreases. A collision-based
analysis shows that $V_{\rm pot} \sim T^{3/2}$: the energy
decreases faster than linearly and vanishes at $T=0$.
Correspondingly, the `excess' energy above the 
hard sphere ground state ($V_{\rm pot}=0$) is small,
explaining the smallness of $\Delta c_V$ at small volume fraction. 
For very dense systems, the energy decreases linearly at low
$T$, and increases with $\phi$ since particles overlap more 
 upon compression. Therefore the larger $\phi$, 
the larger $c_V$ in the fluid phase, and the larger 
is $\Delta c_V$, as observed in fig.~\ref{multi}-d.
These considerations suggest that further analysis of the 
potential enery landscape properties of the present system
and comparison with model landscapes~\cite{heuer2,heuer3} 
could be very valuable.   

In a broad density range 
encompassing $\phi_0$, simulations suggest 
$V_{\rm pot} \sim T^{\mu}$, with $\mu \approx 1.3$, 
a value intermediate between the dilute and dense limits.
Due to the harmonic nature of the potential,
$\sqrt{V_{\rm pot}}$ represents the average overlap between
interacting particles, and suggests a way to quantitatively estimate 
$\phi_{\rm eff}$ in eq.~(\ref{mapping}), and to justify the 
form of the scaling variable in eq.~(\ref{critical}). These
energetic considerations are a suggestive 
physical interpretation, rather than a rigorous
derivation, of the success of eq.~(\ref{mapping}) 
at collapsing data in fig.~\ref{taualpha}-c.

\section{Non-equilibrium jamming at $T=0$}
 
Finally, the low-$T$, non-equilibrium behaviour of the energy density 
during slow annealing is also informative. 
For $\phi < \phi^\star \approx 0.662$, $V_{\rm pot}(\phi,
T \to 0)= 0$, while $V_{\rm pot}$ remains finite above $\phi^\star$. 
Therefore, the nature of the $T=0$ glasses produced 
by slow annealing at volume fractions above point $G$
changes at $\phi^\star$ where a `jamming' transition
similar to the one described in~\cite{pointJ} occurs.
Note that $\phi^\star$ is larger than $\phi_J \approx 0.648$, the critical 
density for jamming determined in~\cite{pointJ} for the same system, 
because our glasses have been annealed--with no trace of 
demixing or crystallization~\cite{torquato2,torquato}.
We emphasize that these jamming transitions
occur at different densities when different protocols 
are used~\cite{torquato},
while the equilibrium glass transition at $\phi_0$ analyzed in this work 
is uniquely defined by the behaviour of the equilibrium 
relaxation time, so that both transitions should be considered as
distinct phenomena~\cite{jorge}.

\section{Discussion and conclusion}

The dynamic scaling behaviour in eq.~(\ref{critical})
describing the interplay between density and temperature 
for soft repulsive particles is in stark contrast with the 
results obtained for soft spheres interacting with a pure 
inverse power law potential~\cite{casalini,alba}, because
the interaction potential of elastic spheres has a finite cutoff
(the particle diameter). Therefore, elastic spheres reduce 
to hard spheres in the small-$T$ limit, which 
is crucial to obtain a change in fragility. We believe
our results would generically carry out for repulsive 
potentials with a finite range. 

A second interesting feature of this study 
is that the location of a divergence for the equilibrium 
relaxation time is extrapolated from a demanding scaling procedure
where a large set of independent data is used to locate a 
critical density. To the best of our knowledge, 
such an analysis using two control parameters has
no counterpart in the glass transition literature. In particular, 
we believe it yields a rather accurate determination 
of the critical density $\phi_0$ where the equilibrium 
relaxation time of the hard sphere fluid extrapolates 
to $\infty$, but leaves open the existence of a finite 
temperature singularity above $\phi_0$. 
Our determination of the existence of a genuine 
divergence of $\tau_\alpha$ for hard spheres at $\phi_0$
is based on the analysis of the metastable fluid--demixing 
and crystallization being avoided 
due to polydispersity. Whether this ideal 
glass transition is itself avoided~\cite{torquato} due 
to crystallization or demixing at much larger density 
is another important issue that our data leave open.

In conclusion, our preliminary studies 
of repulsive elastic particles at thermal equilibrium 
suggest that soft repulsive particles 
are a promising new tool to gain deeper understanding of 
glass transition phenomena in colloidal and molecular systems. 
Experimentally, this could be
directly realized using soft colloidal particles.

\acknowledgments

Fruitful exchanges with 
G. Biroli,
P. Chaudhuri,
L. Cipelletti,
W. Kob,
J. Kurchan,
S. Nagel,
D. Reichman, and
G. Tarjus
are acknowledged.
L. B. thanks the University of Chicago and Argonne National Laboratory
for financial support in early stages of this work in 2007.

\end{document}